# Self-heterodyne spectroscopy via a non-uniformly spaced frequency comb

Bofeng Zhang[1,2†], Gang Zhao[1,2*†], Xiaobin Zhou[1,2], Xiaojuan Yan[1,2], Jiaqi Yang[1,2], Weiguang Ma[1,2*], Suotang Jia[1,2]

[1]State Key Laboratory of Quantum Optics Technologies and Devices, Institute of Laser Spectroscopy; Shanxi University, Taiyuan, 030006, China.
[2]Collaborative Innovation Center of Extreme Optics; Shanxi University, Taiyuan, 030006, China.
*Corresponding author. Email: gangzhao@sxu.edu.cn (Gang Zhao); mwg@sxu.edu.cn (Weiguang Ma)
†These authors contributed equally to this work: Bofeng Zhang, Gang Zhao.

## Abstract

Frequency comb spectroscopy has significantly advanced molecular spectroscopy across scientific research and diverse applications. Among its key performance metrics especially for time-resolved studies, sensitivity and measurement speed are paramount. However, a long-standing compromise between these parameters arises from the need for noise reduction. Here, we introduce a comb spectroscopy system that overcomes this limitation using a single frequency comb of non-uniformly spaced modes. The comb is generated using an extremely simple setup, composed of a continuous-wave (CW) fiber laser and a single-sideband phase modulator (SSM). Our approach delivers optical-to-radio-frequency conversion comparable to dual-comb spectroscopy (DCS) but through a simplified self-heterodyning architecture. By leveraging the intrinsic mutual coherence of the comb, this design achieves a noise-equivalent absorption coefficient (NEA) of $5.0\times10^{-6}$ $Hz^{-1/2}$—an order-of-magnitude improvement over state-of-the-art DCS, coupled with long-term stability. The system resolves weak molecular overtone spectra on nanosecond timescales, in a single-shot measurement, at a signal-to-noise ratio of 128. This integration of high sensitivity, resolution, and speed resolves the core trade-off that has long constrained time-resolved spectroscopic analysis.

## Introduction

Frequency comb spectroscopy leverages a series of laser modes to interrogate molecular spectra and serves as a powerful analytical tool for applications spanning from tests of fundamental physical laws to trace gas sensing[1–7]. Extending its utility to applications requiring high-speed detection of weak molecular absorption—such as chemical reaction kinetics and combustion diagnostics[8–11]—necessitates the development of next-generation comb spectroscopy systems that combine high sensitivity with rapid response. Among various comb spectroscopies, dual-comb spectroscopy (DCS) has emerged as a particularly powerful approach[12–14], especially well-suited for time-resolved measurements with high spectral resolution[15–18]. This approach leverages the multiheterodyne of two combs, enabling comb-mode-resolved measurements at high acquisition rates without any moving components[19–24].

Despite these advantages, the full potential of DCS is constrained by its underutilized detection sensitivity. While direct absorption spectroscopy (DAS) has achieved a noise-equivalent absorption coefficient (NEA) of $10^{-5}$ $Hz^{-1/2}$ with microwatt-level laser power[25], reported sensitivities of DCS have yet to approach this benchmark[26], even when operating at high power[27,28]. A key obstacle to bridging this performance gap is the mutual incoherence inherent in the dual comb architecture. Existing mitigation strategies, including active feedback[29], adaptive sampling[30,31], and computational methods[32–34], necessitate a low repetition rate difference[35,36] or multi-cycle acquisition. These requirements inevitably reinforce the unresolved trade-off between sensitivity and measurement time—a pervasive limitation in absorption spectroscopy[37].

Here, we introduce a frequency comb spectroscopy scheme that overcomes the sensitivity-speed trade-off. Our system employs a single frequency comb with engineered, non-uniform mode spacing. Analogous to DCS, it enables optical-to-radio-frequency conversion, but accomplishes this through a simplified, self-heterodyning architecture that requires no second comb. Leveraging the intrinsically high coherence between the teeth, this approach yields a signal-to-noise ratio (SNR) of $2.0\times10^5$ $Hz^{1/2}$ for beat notes, enabling spectral acquisition on nanosecond timescales. This results in a demonstrated sensitivity of $5.0\times10^{-6}$ $Hz^{-1/2}$ for molecular spectra, coupled with long-term stability over 400 seconds, all without requiring phase compensation. The proposed comb, comprising only a continuous-wave (CW) laser and a single-sideband phase modulator (SSM), is notably simple, compact, and robust. We explain its principle, detail the setup, quantify its spectroscopic sensitivity, and demonstrate applications in time-resolved measurement and multi-gas detection.

## Results

### Basic principle

The operating principle of our self-heterodyne spectroscopy, which utilizes a single non-uniform frequency comb, is illustrated in Fig. 1a. After interaction with gas medium, the comb is detected, and its spectral information is down-converted to the radio frequency (RF) domain via a self-heterodyne process. This yields an RF comb via fast Fourier transform (FFT), where each line corresponds to the difference frequency between adjacent optical teeth, faithfully encoding the molecular absorption features. A fundamental distinction from DCS[38–40] (illustrated in Fig. 1b) lies in the signal origin: Since all heterodyne beats derive from a single comb, inherent mutual coherence is guaranteed, producing a high-fidelity RF comb free from decoherence noise.

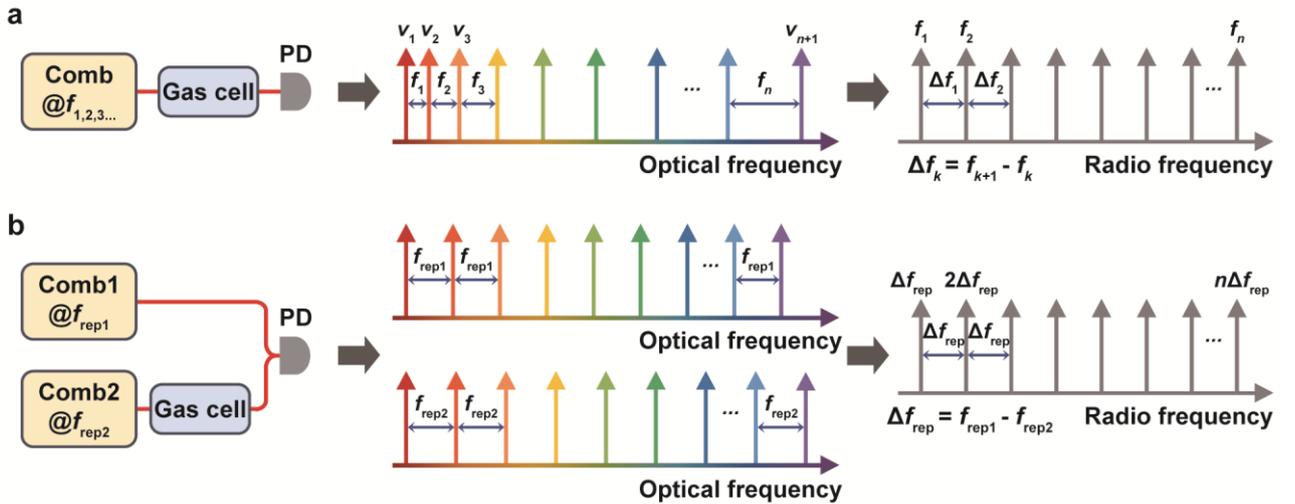

**Fig. 1. Concept of multiheterodyne spectroscopy using frequency comb. a** Schematic of a non-uniform comb spectroscopy. A frequency comb with unequally spaced lines has frequency interval of $f_k = f_1, f_2, …, f_n$. The beats of the adjacent lines produce an radio frequency (RF) comb with frequencies corresponding to $\Delta f_k = f_{k+1} - f_k$. **b** Dual-comb spectroscopy (DCS). Two combs with slightly different repetition rates, $f_{rep1}$ and $f_{rep2}$, beat against each other on a photodetector (PD), yielding a series of RF lines.

The proposed comb can be represented by a superposition of an arbitrary set of $n+1$ discrete optical fields in the time domain, which can be expressed as $E_0 = \sum_{k=1}^{n+1} A_k \exp[i(2\pi v_k t + \phi_k)]$, where $k = 1, 2, …, n+1$ represents the index of these discrete optical fields with respective frequencies $v_k$, and $\phi_k$ denotes their initial phase. The frequency interval between adjacent fields is defined as $f_k = v_{k+1} - v_k$. To avoid frequency interference in the RF domain, the comb's frequency distribution must ensure that all frequency intervals between the comb teeth are unique. This is guaranteed by two constraints: (1) the spacings between adjacent optical fields are non-uniform: $f_i \neq f_j$ for all $i \neq j$; (2) the interval between any pair of adjacent optical fields differs from that between any pair of non-adjacent fields: $f_k \neq v_i - v_j$ for all $i - j \geq 2$.

Although the values of $f_k$ can be set arbitrarily provided they satisfy the two constraints, it is advantageous to set the intervals $\Delta f_k$ (equal to $f_{k+1} - f_k$) to a constant value $\Delta f$ and to set $f_1$ to an integral ($m$) multiple of $\Delta f$ (see Supplementary Note 1). This configuration allows the entire comb to be fully resolved within a single measurement period of $1/\Delta f$, corresponding to one period of the time-domain signal, which facilitates the kinetic measurement. In this scenario, the modulation frequency intervals can be expressed as $f_k = f_1 + (k - 1)\Delta f = (m + k - 1)\Delta f$.

After transmission through a gas medium, the beat notes, generated at $f_k$ from adjacent comb teeth, can be expressed as

$$I_\alpha^{f_k} = 2\sum_{k=1}^{n} A_k A_{k+1} \exp\left\{-\left[\alpha(v_k) + \alpha(v_{k+1})\right]/2\right\} \cos\left(2\pi f_k t + \phi_{k+1} - \phi_k + \varphi_k - \varphi_{k+1}\right) \quad (1)$$

where $\alpha(v_k)$ is the frequency-dependent absorption coefficient, and $\varphi(v_k)$ is the phase shift introduced by the gas. By performing a balanced detection between the beat notes with and without absorption, then applying a FFT, the attenuation of the RF comb caused by gas absorption is quantified as:

$$R(k) = \exp\left\{-\left[\alpha(v_k) + \alpha(v_{k+1})\right]/2\right\} \quad (2)$$

This establishes a direct relationship between the RF comb and the target gas absorption information, thereby providing a method for quantitative gas concentration sensing. By fitting the measured data using the model in Eq. (2), the absorption coefficient and gas concentration can be accurately derived.

As a key parameter for assessing coherence in DCS, the linewidth of the beat notes in our system is governed solely by the phase noise of the modulation frequencies, which is significantly narrower than that achieved in DCS (See Supplementary Note 2). This indicates that our new spectrometer can provide superior spectral resolution. Remarkably, even when using a free-running CW laser, the coherence between the teeth is maintained. This capability is invaluable for high-resolution spectroscopy, especially for retrieving the broadening coefficients of molecular transitions[41,42] or observing Doppler-free spectroscopy with narrow spectral width[43,44].

**Experimental setup**

The experimental realization of our single-comb spectrometer is depicted in Fig. 2a. The setup is driven by a CW fiber laser at 1.53 μm. The laser light is sent to a fiber-coupled SSM, a key deviation from the conventional electro-optic modulator (EOM) typically used in modulation based DCS. This SSM is driven by an arbitrary waveform generator (AWG), which is programmed to output repetitive time-domain signal corresponding to an RF comb of unequally spaced frequencies. Through the SSM, the carrier and upper sidebands are suppressed, leaving only the lower sidebands to prevent frequency aliasing, thus generating the final optical frequency comb with non-uniform mode spacing. The resulting comb is split by a 1:1 fiber splitter into two paths: a reference beam, detected directly by photodetector PD1, and a probe beam, which passes through a gas cell containing acetylene before being measured by PD2. The optical power at both photodetectors is maintained at approximately 20 μW. The beat signals from the photodetectors are captured by a digitizer operating at 1 GS/s and are processed via FFT to generate the RF comb, from which the gas absorption information is extracted by spectral fitting.

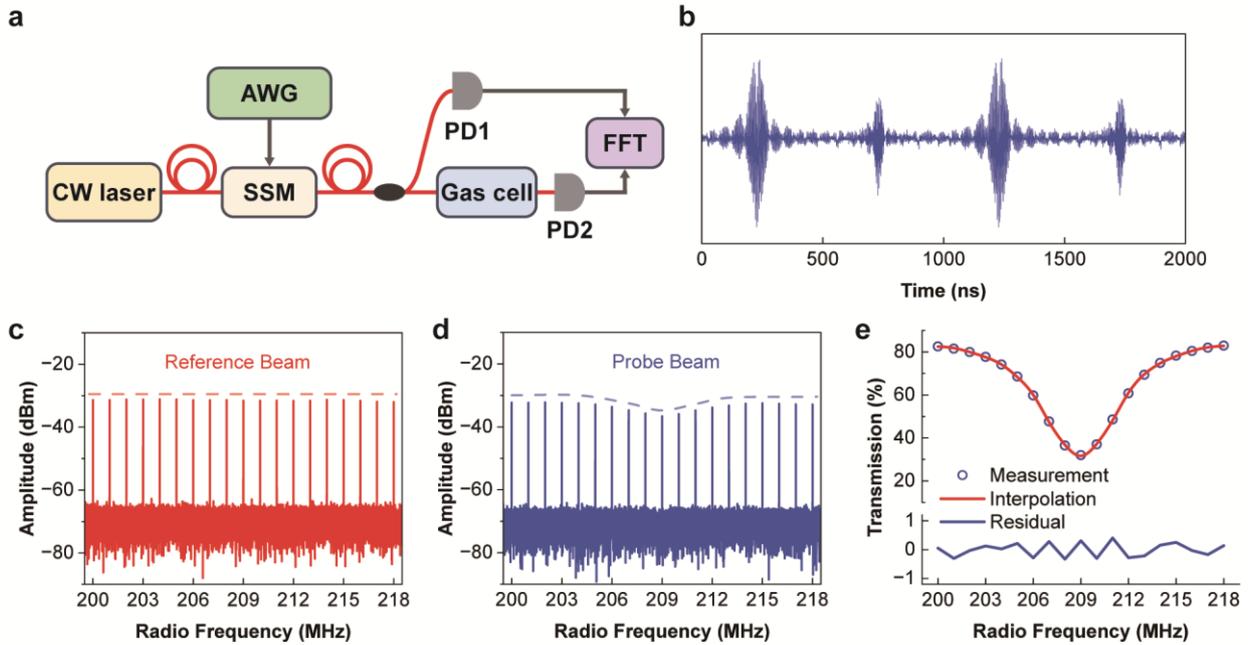

**Fig. 2. Experimental setup of the non-uniformly spaced frequency comb and corresponding spectroscopy measurements. a** The comb is generated by phase-modulating a continuous-wave (CW) laser. An arbitrary waveform generator (AWG) drives a single-sideband phase modulator (SSM) using a signal composed of a series of unequally spaced radio frequency (RF) lines in the frequency domain. The resulting optical frequency comb is then split into two parts, which serve as reference and probe light, respectively. The outputs from two photodetectors, PD1 and PD2, are analyzed using a fast Fourier transform (FFT) to produce frequency spectra. **b** Time-domain interferogram of the reference signal, spanning a duration of 2 μs. **c** FFT spectra of the interferograms acquired over 1 ms for the reference beam. **d** FFT spectra of the interferograms acquired over 1 ms for the probe beam. **e** Retrieved molecular spectral (blue dots) and its fitting result (red curve) with speed-dependent Voigt profile. The lower panel depicts the fitting residual.

To minimize the excitation of high-order and spurious frequency components that can cause crosstalk with the 1st sidebands (See Supplementary Note 3), the modulation index is maintained at a value much less than unity. As a result, the spectral coverage of our comb is dictated by the bandwidth of the SSM, which is 17 GHz in our implementation. In our setup, the number of teeth and their corresponding frequencies are directly programmed via the modulation frequencies applied by the AWG, making the system highly adaptable to diverse spectral resolution requirements.

**Characterization of spectroscopic measurement**

We next demonstrate the spectroscopic capabilities of our system. In this proof-of-concept experiment, a total of 20 comb teeth are generated within the system. The modulation frequencies are configured with $\Delta f = 1$ MHz and $f_1 = 200$ MHz, plus a 1 GHz offset. This yields an ascending series from 1000 to 4971 MHz. This configuration provides a spectral coverage of approximately 4 GHz with a resolution of around 200 MHz, which is sufficient to resolve molecular lines under Doppler-broadened conditions. The modulation index for each frequency component was maintained at 0.065. The detected time-domain beat notes between adjacent comb teeth, shown in Fig. 2b, exhibit a period of 1 μs. A FFT of these signals converts them into an RF comb, whose spectrum for both the reference and probe beams (acquired over 1 ms) is displayed in Fig. 2c and 2d, respectively. The RF comb exhibits frequencies $f_k$ ranging from 200 MHz to 218 MHz with a uniform spacing of $\Delta f = 1$ MHz. Both spectra reveal nineteen distinct peaks, each corresponding to the heterodyne beat between a pair of adjacent optical comb teeth. The dips evident in the probe spectrum are attributed to gas absorption. The retrieved molecular spectrum is plotted as dots in Fig. 2e. Spectral fitting is performed using a model based on Eq. (2), which incorporates crosstalk (Supplementary Note 3), electric field modulation by a phase modulator and a speed-dependent Voigt profile. The spectral parameters for the $C_2H_2$ transition at 6531.78 cm$^{-1}$ are taken from the HITRAN database[45]. The fitted spectrum (red curve) agrees well with the data, as evidenced by the residuals in the lower panel, yielding a spectral SNR of 213.

To evaluate the detection sensitivity, we conduct a spectral SNR analysis on individual RF teeth and an Allan-Werle analysis. As shown in Fig. 3a, the SNR scales linearly with the number of acquired interferograms up to $10^6$, with a measured slope of $2.0 \times 10^5$ Hz$^{1/2}$. This slope is superior to those reported for state-of-the-art DCS systems, as detailed in Table S1. For the Allan-Werle analysis, molecular absorption spectra of a single sample are measured consecutively over an extended time. Our maximum acquisition rate is determined by the frequency spacing of the RF comb, which here is 1 MHz. In practice, however, the digitizer's onboard memory limits single-burst acquisition to 4000 consecutive interferograms, corresponding to a total measurement time of 4 ms. Additionally, due to the dwell time between acquisitions, the effective data acquisition rate for long-term measurements is limited to 3.5 Hz. Consequently, two Allan-Werle plots of absorption coefficient are presented in Fig. 3b: the blue curve corresponds to the 4 ms measurement at 1 MHz rate, while the black curve represents the long-term measurement over 2500 s at 3.5 Hz. At the high rate, a NEA at 1 s of $5.0 \times 10^{-6}$ Hz$^{-1/2}$ is achieved even with a modest comb tooth power of 1 µW. This performance enables the detection of absorptions as weak as $5.0 \times 10^{-3}$ with a 1 µs time resolution. The long-term stability is reflected in the continual decrease of the detectable absorption coefficient with integration time, reaching a sensitivity of $6.1 \times 10^{-6}$ at an integration time of 442 s, even without fully exploiting the acquisition capability. The observed long-term instability is attributed to power drift of the AWG and non-linearity of the acquisition system.

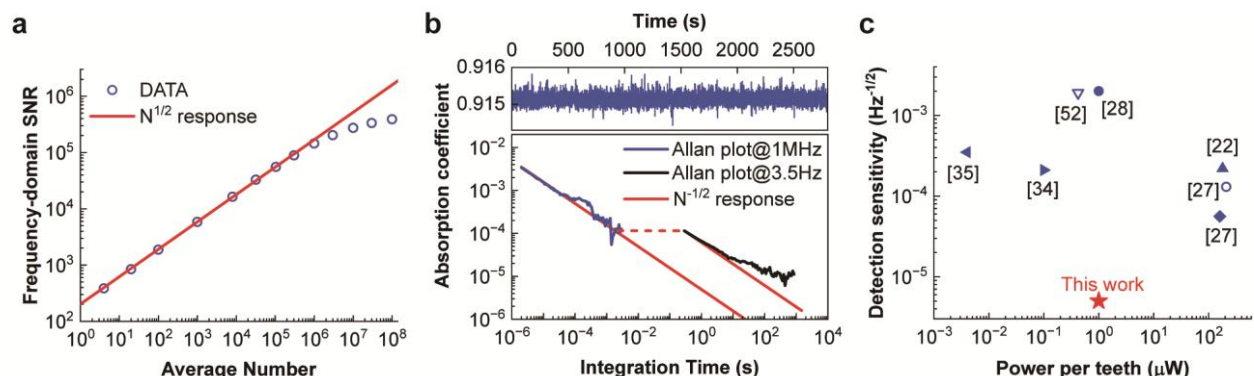

**Fig. 3. Evaluation of the detection sensitivity. a** frequency-domain signal-to-noise ratio (SNR) (blue dots) as a function of interferogram number. The red line indicates a linear behavior. **b** Analysis of detection sensitivity for molecular spectroscopy using the Allan-Werle analysis. The upper panel shows the retrieved absorption coefficient from consecutive measurements of a single sample over 2800 s at a rate of 3.5 Hz. The black curve in the lower panel presents its Allan-Werle plot, where the red curve represents a white-noise response. And the blue curve in the panel demonstrates the result when the spectral acquisition capability is fully utilized at an interferogram acquisition rate of 1 MHz. **c** comparison of detection sensitivity between our spectrometer and state-of-the-art DCS systems.

Fig. 3c benchmarks the detection sensitivity of our spectrometer against state-of-the-art DCS systems, with a comprehensive summary provided in Supplementary Note 4. Sensitivity values are derived from the Allan-Werle plot or spectral fitting residuals. Whereas existing DCS implementations typically achieve sensitivity on the order of $10^{-4}$ Hz$^{-1/2}$, with the best value reported to date being, to our knowledge, of $5.6 \times 10^{-5}$ Hz$^{-1/2}$ at 160 µW per tooth[27]. In contrast, our architecture attains a sensitivity of $5.0 \times 10^{-6}$ Hz$^{-1/2}$ using only 1 µW per tooth. This represents an order-of-magnitude improvement, achieved with a markedly simplified optical setup and without any phase compensation techniques. Moreover, our performance also surpasses the state-of-the-art sensitivity levels in DAS, which is on the order of $10^{-5}$ Hz$^{-1/2}$ in the sub-milliwatt power regime[25]. These results collectively underscore the inherent mutual coherence of our single comb source. It is worth noting that instead of assessing the long-term stability via SNR of the radio spectrum, we prefer to use Allan-Werle analysis for the absorption coefficient. This approach is favored because the former method often fails to adequately account for signal fluctuations.

The fundamental sensitivity limit of our system is determined by the shot noise, given by (see Supplementary Note 5):

$$\alpha_{\min} = \sqrt{\frac{2e(n+1)\Delta f_b}{\eta_c P_0}} \qquad (3)$$

where e is the elementary charge, $n+1$ is the total number of optical frequency teeth, $\Delta f_b$ is the detection bandwidth, $\eta_c$ is the photodiode responsivity (A/W), and $P_0$ is the incident optical power per comb tooth in the absence of absorption. Under our experimental conditions, the calculated shot-noise limited sensitivity is $8.4\times10^{-7}$ Hz$^{-1/2}$. The measured white noise response of our system ($5.0\times10^{-6}$ Hz$^{-1/2}$) lies a factor of 5.9 above the shot noise limit. This discrepancy is primarily attributed to noise contribution from the detection chain, including the photodetector and data acquisition electronics.

**Time-resolved measurement**

We then employed our spectrometer for kinetic studies of molecular spectra to demonstrate its time-resolved measurement capability. In conventional DCS implementations, achieving microsecond-scale time resolution commonly necessitates the use of combs with repetition rates exceeding GHz[46,47]. These high repetition rates are required to enable multi-cycle measurement or signal averaging essential for maintaining adequate sensitivity. However, this operational paradigm inherently limits spectral resolution, rendering such systems unsuitable for resolving molecular lines under Doppler-broadened conditions at low pressure.

In contrast, the high mutual coherence of our single comb source provides intrinsic sensitivity that eliminates the need for multi-cycle measurement or signal averaging, thereby enabling direct high-speed spectral acquisition. To fully harness this potential, we reconfigured the AWG with $f_1$ = 400 MHz and $\Delta f$ = 50 MHz, thereby generating a new comb with inter-tooth spacing ranging from 400 to 750 MHz. This configuration produces a single-interferogram time-domain period of 20 ns, as shown in Fig. 4a. A high-speed photodetector and oscilloscope, each with a bandwidth exceeding 1 GHz, were employed to accurately capture the transient signal. The molecular spectrum retrieved from a single 20 ns acquisition is presented in Fig. 4b (blue dots). The corresponding fit, depicted by the red curve, shows good agreement with the data and yields a SNR of 128. The current system resolves up to 9 comb teeth per spectrum, a constraint imposed by the finite detection bandwidth and the need to mitigate crosstalk-induced distortion. We emphasize that both the number of resolvable teeth and the acquisition rate can be directly scaled by utilizing a photodetector with broader bandwidth.

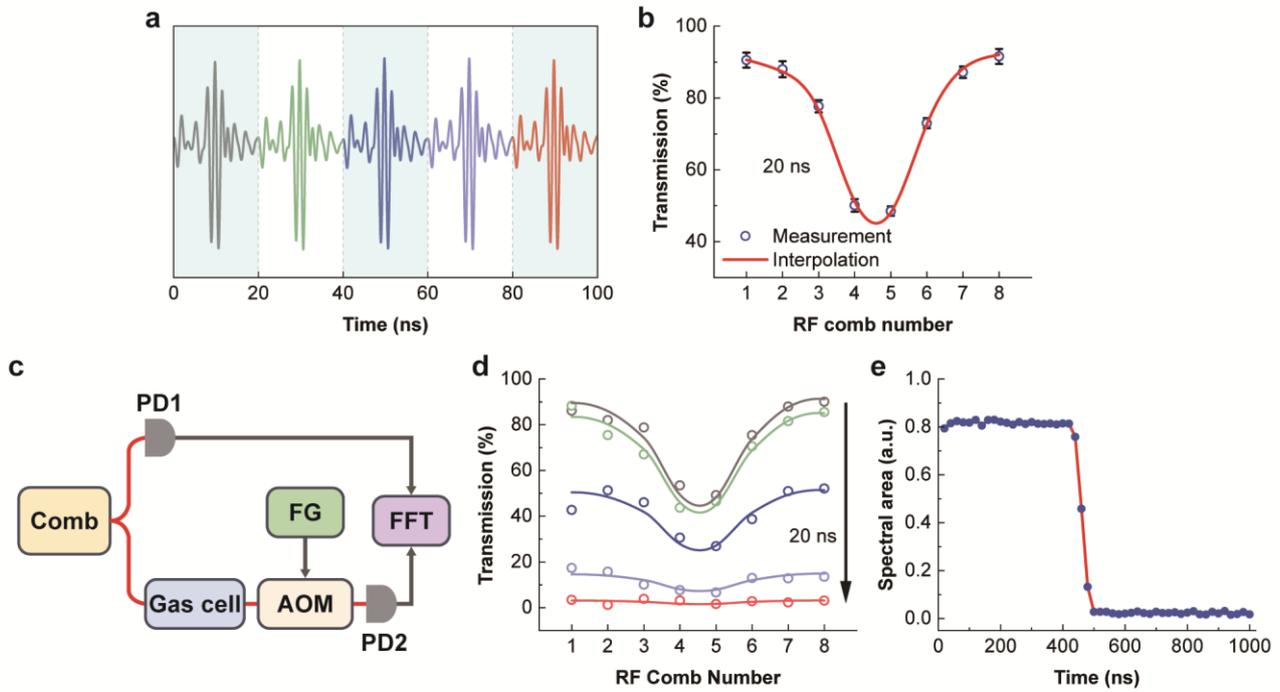

**Fig. 4. Time-resolved measurement with a 20-ns resolution. a** Time-domain interferogram of the reference signal with a period of 20 ns. **b** Molecular absorption spectra acquired at 20 ns, along with its fitting results. The error bar is the standard deviation of the magnitude of individual measured teeth. **c** Configuration of the experimental setup for high-speed verification of time-resolved measurements. **d** Absorption spectrum obtained through sequential measurement. **e** Measurement of rapid changes in integrated absorption area with a time interval of 20 ns.

To further verify the high acquisition rate, we adopt an approach similar to ref. 16, using an acoustic-optic modulator (AOM) placed in the probe beam path to rapidly switch off the light and thereby simulate a transient change in gas concentration (Fig. 4c). The integrated area under each spectrum serves as a proxy for the instantaneous concentration. The time resolved retrieved spectra at 20 ns resolution are shown in Fig. 4d, and the corresponding retrieved integrated area are shown as blue dots in Fig. 4e. The resulting curve exhibits a clearing falling edge with a slope that matches the AOM's characteristic fall time of 30 ns.

Notably, our approach achieves an acquisition speed comparable to ref. 16, while simultaneously delivering higher spectral fidelity and a tenfold improvement in frequency resolution, despite operating at optical power levels more than 300 times lower. Together, these advances establish a practical and powerful platform for applying frequency combs to time-resolved spectroscopy.

**Multigas detection**

We further demonstrate the versatility of our platform by extending its application to multigas detection. To increase the spectral coverage, we integrated a second frequency comb, exploiting the ability to assign distinct, non-overlapping modulation frequencies to each comb. A new comb is generated using a 1583 nm CW laser source and a separate SSM, configured with $\Delta f$ = 1MHz and $f_1$ = 232 MHz, and a 6 GHz offset. This configuration produces a unique set of frequencies from 6000 to 10579 MHz, with the corresponding RF comb spanning from 232 MHz to 250 MHz, thereby ensuring clear spectral separation from the first comb. As illustrated in Fig. 5a, the two combs, operating in complementary spectral bands, are combined to simultaneously measure $C_2H_2$ and $CO_2$ gases. The combined time-domain signal (Fig. 5b) maintains a 1 μs period. The resulting RF spectra (Fig. 5c) clearly resolve the absorption lines of the two different gases within their respective bands. The retrieved absorption spectra (Fig. 5d) exhibit a SNR comparable to that achieved using a single comb. This approach provides a practical and scalable route to expand frequency coverage—a capability that is challenging to realize with conventional DCS.

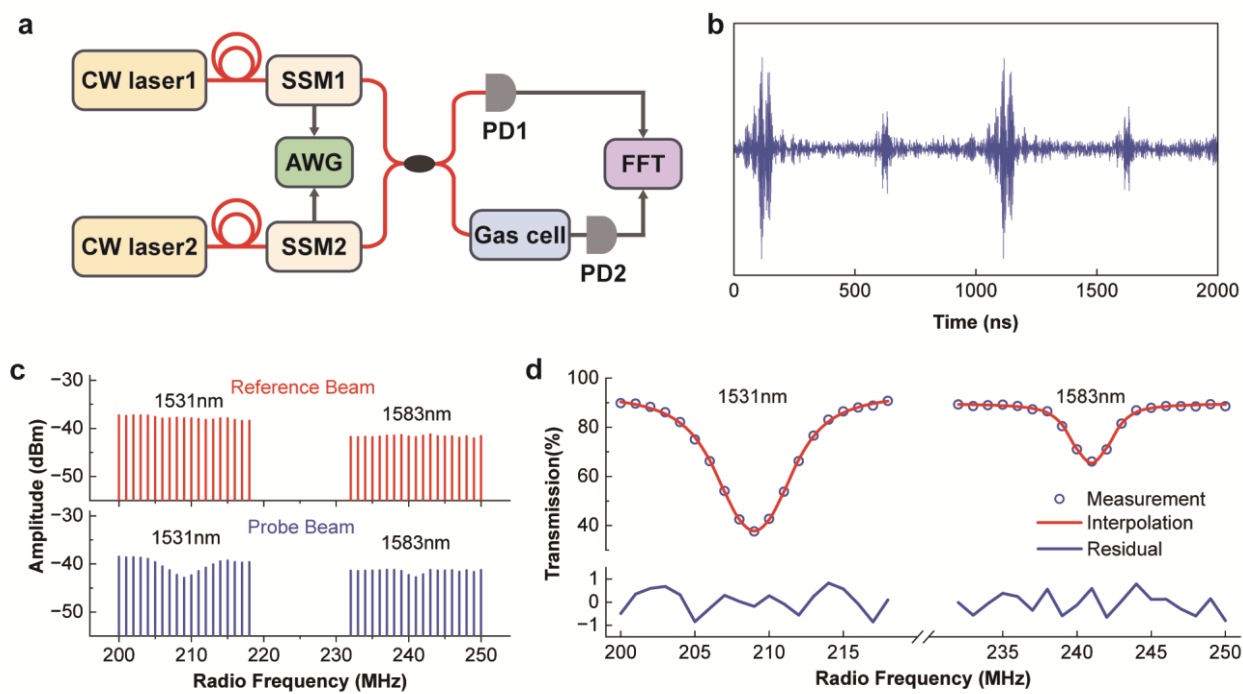

**Fig. 5. Multigas detection through spectral coverage extension. a** A spectrometer for multigas detection based on the combination of two non-uniform combs. **b** Time-domain interferogram of the reference signal, spanning a duration of 2 μs. **c** Radio frequency (RF) comb spectra of the reference (upper panel) and probe (lower panel) beam, composed of two clusters of comb teeth centered at 1531 nm and 1583 nm, respectively. The absorption features of $C_2H_2$ and $CO_2$ are evident as dips within each cluster. **d** Retrieved absorption spectra and their fitting results. The lower panel depicts the fitting residuals.

## Discussion

Here, we develop a comb-spectroscopic technique that overcomes the sensitivity–measurement time trade-off by utilizing a non-uniform frequency comb. The key advantages include: (1) high-speed spectral acquisition, (2) an order-of-magnitude improvement in sensitivity over state-of-the-art DCS, (3) high-resolution spectroscopic capability, (4) a simple and compact architecture conducive to field deployment. These features collectively render the technique well suited for time-resolved spectroscopy.

This technique also holds strong potential for remote atmospheric sensing. Its high acquisition rate can effectively mitigate signal degradation from atmospheric turbulence[48]. Moreover, by equalizing the comb tooth amplitudes, spectra could be retrieved directly from transmitted light, eliminating two critical limitations of conventional DCS for remote applications[49]: the need for frequency reference distribution between transceiver units[50], and the mandatory phase stabilization.

In summary, this technique establishes a paradigm for high-performance molecular spectroscopy, paving the way for compact, high-sensitivity instruments capable of time-resolved kinetic studies and atmospheric monitoring, even in challenging environments.

While the present study focuses on improving properties in terms of sensitivity and acquisition rate, we recognize that spectral coverage in its current form remain limited. Future efforts will prioritize bandwidth expansion through methods such as cascading modulators[51] or coherently combining multiple combs at different wavelengths—an approach uniquely suited to our architecture, as preliminarily demonstrated in Fig. 5. Extending operation into the mid-infrared via difference-frequency generation[52] is a compelling path to accessing stronger molecular fundamental bands. Additional sensitivity gains could be realized by increasing optical power or leveraging resonant enhancement schemes[53–57].

## Methods

### Generation of the non-uniform frequency comb

The non-uniform frequency comb was generated using the following experimental procedure. First, a custom program is used to generate the time-domain radio frequency (RF) signal data, which comprises 20 unequally spaced frequency components as described in the main text. This signal is output by an arbitrary waveform generator (AWG, Tektronix AWG7002A), amplified by a power amplifier (Mini-Circuits ZHL-1W-63-S+), and then fed into a fiber-coupled single-sideband phase modulator (SSM, Keyang Photonics KY-SSB-15-0118). The SSM operates in the optical frequency range of 1.525 μm to 1.610 μm with a modulation bandwidth of 17 GHz and incorporates an internal feedback loop to actively suppress the optical carrier and the unwanted sideband. Prior to the SSM, a 10 dB RF attenuator (Mini-Circuits VAT-10W2+) is inserted to adjust the power of each RF component to –7 dBm, maintaining the modulation index per comb tooth at approximately 0.065. This low modulation index effectively mitigates the impact of RF power drift on long-term system stability. The optical source is a continuous-wave (CW) fiber laser (NKT Photonics Koheras Adjustik E15 PztS PM) operating at 1531 nm with an output power of about 27 mW. To prevent saturation or nonlinear response in the photodetector and data acquisition system, the optical signal was further attenuated by a 3 dB fiber attenuator (Lbtek FFA-03S-APC). This ensured that the input power to the subsequent 1:1 fiber splitter (Flyin) and the photodetector (PD, Micro Photonics BAPD-600M-A) was maintained at about 20 μW.

### Detection and data acquisition system

The probe light is directed through a 5.5 cm long gas cell filled with pure $C_2H_2$ at a pressure of approximately 55 Torr. For $CO_2$ measurements, another laser operated at 1583 nm and a multipass cell with an effective path length of 10 m are used. The transmitted light is detected by a photodetector, and the resulting electrical signal is acquired by a data acquisition system (GaGe PMX-161-G20) operating at a sampling rate of 1 GS/s with 16-bit resolution. Impulse noise, which is suspected to originate from SSM feedback-loop instability, is effectively suppressed by digital filtering in post-processing.

To mitigate the potential impact of laser carrier frequency drift on measurement accuracy (See Supplementary Note 6), an additional feedback control loop is implemented to actively stabilize the laser frequency. This loop generates an error signal based on the symmetry of the amplitudes of the two comb teeth flanking the absorption peak, which is then fed back to the piezoelectric transducer (PZT) of the laser cavity. This active stabilization enhances the system's long-term stability, ensuring reliable and precise spectral measurements. Further improvement in frequency stability could be achieved by phase-locking the laser carrier to an atomic clock-referenced frequency comb.

### Experimental conditions for time-resolved measurements

For the time-resolved measurements, the modulation frequencies are adjusted, resulting RF spectral lines spanning 400 MHz to 750 MHz. This wider spectrum necessitates a larger detection bandwidth. To meet this requirement, we employed two 1 GHz-bandwidth photodetectors (Newport 1611FC-AC) and a 4 GHz-bandwidth oscilloscope (Rohde & Schwarz RTO64). Due to the detectors' lower responsivity and higher saturation power, the optical power delivered to each detector was increased to 1.6 mW, ensuring approximately 200 μW of available power per optical comb tooth.

**Acknowledgments**

The authors acknowledge funding from the National Natural Science Foundation of China (61905136, 62175139), the Key R&D Program of Shanxi Province (202302090301014), the Central Guidance on Local Science and Technology Development Fund of Shanxi Province (YDZJSX2024D001), and the Fundamental Research Program of Shanxi Province (202403021222027).